# Gas equilibration gas modulation refractometry (GEq-GAMOR) for assessment of pressure with sub-ppm precision

Isak Silander, Thomas Hausmaninger, Clayton Forssén,

Department of Physics, Umeå University, SE-901 87 Umeå, Sweden

Martin Zelan [a],

Measurement Science and Technology, RISE Research Institutes of Sweden, SE-501 15 Borås, Sweden

Ove Axner [b]

Department of Physics, Umeå University, SE-901 87 Umeå, Sweden

[a] Electronic mail: martin.zelan@ri.se
[b] Electronic mail: ove.axner@umu.se

Gas Modulation Refractometry (GAMOR) has recently been developed to mitigate drifts in the length of the cavities in Dual-Fabry-Perot Cavity (DFPC) based refractometry. By performing repeated reference assessments with the measurement cavity being evacuated while the reference cavity is held at a constant pressure, the methodology can reduce the influence of the long-term drifts, allowing it to benefit from the high precision of DFPC-based refractometry at short time scales. A novel realization of GAMOR, referred to as Gas Equilibration GAMOR (GEq-GAMOR), that outperforms the original realization of GAMOR, here referred to as Single Cavity Modulated GAMOR (SCM-GAMOR), is presented. It is based upon the fact that the reference measurements are carried out by equalizing the pressures in the two cavities. By this, the time it takes to reach adequate conditions for the reference measurements has been reduced. This implies that a larger fraction of the measurement cycle can be devoted to data acquisition, which reduces white noise and improves on its short-term characteristics. The presented realization also encompasses a new cavity design with improved temperature stabilization and assessment. This has contributed to an improved long-term characteristics of the GAMOR methodology. The system was characterized with respect to a dead weight piston gauge. It was found that for short integration times (up to 10 min) it can provide a response that exceeds that of the original SCM-GAMOR system by a factor of two. For integration times longer than this, and up to 18 hours, the system shows, for a pressure of 4303 Pa, an integration time independent Allan deviation of 1 mPa (corresponding to a precision, defined as twice the Allan deviation, of 0.5 ppm). This implies that the novel system shows a significant improvement with respect to the original realization of GAMOR for all integration times (by a factor of 8 for an integration time of 18 hours). When used for low pressures, it can provide a precision in the sub-mPa region; for the case with an evacuated measurement cavity, the system provided, up to *ca.* 40 measurement cycles (*ca.* 1.5 hours), a white-noise limited noise of 0.7 mPa $(cycle)^{1/2}$, and minimum Allan deviation of 0.15 mPa. Furthermore, over the pressure range investigated, i.e. in the 2.8 – 10.1 kPa range, it shows, with respect to a dead weight piston gauge, a purely linear response. This implies that the system can be used for transfer of calibration over large pressure ranges with exceptional low uncertainty.

# I. INTRODUCTION

Fabry-Perot cavity (FPC) based refractometry is a sensitive technique for assessment of gas refractivity, density, pressure, and gas flows.[1-17] The technique is built on the principle that the frequency of a cavity mode in an FPC is shifted when gas with a given refractivity is let into the cavity.[1-3, 15] Recent works have indicated that the technique has the potential to replace current pressure standards, in particular in the 1 to 100 kPa range.[18-24] With the revision of the SI-system in May 2019, in which the Boltzmann constant will be defined as a fixed value without any uncertainty, the performance would, in principle, be limited only by the accuracies of quantum calculations of gas parameters and the determination of the gas temperature.

Ordinary FPC refractometry is in general limited by the stability of the length of the cavity;[17, 25-27] a length change of 1 pm of a 30 cm long cavity during a measurement corresponds, for $N_2$, to an uncertainty in the assessment of pressure of 1 mPa. To achieve this degree of precision with FPC-based refractometry an exceptional mechanical stability is required.

A means to alleviate this is to utilize a dual FP cavity (DFPC) in which two cavities are bored in the same spacer block; in this case one cavity serves as the measurement cavity (in which gas is let in and pumped out) while the other is the reference cavity.[3, 13-18, 26, 27] For each cavity, a laser is locked to one of the cavity modes and the beat frequency of the two lasers is measured. Any change in length of the cavity spacer that is common to the two cavities will then cancel in the beat frequency and not affect the assessment.

Despite this, since the two cavities in a DFPC set-up can drift dissimilarly, the highest performance still requires extraordinary stable conditions. As a means to remedy this, Gas Modulation Refractometry (GAMOR) has recently been developed.[28, 29] By performing repeated reference assessments with the measurement cavity being evacuated (while the reference cavity is held at a constant pressure), the methodology can significantly reduce the influence of the long-term drifts (i.e. eliminating its linear parts) that are mainly caused by length changes of individual cavities. This allows it to benefit from the high precision FPC-based refractometry has at short time scales also at long time scale. It was recently shown that a GAMOR instrumentation with a non-temperature-stabilized cavity spacer, could, when referenced to a dead weight piston gauge, reduce the influence of drifts of the cavities three orders of magnitude with respect to a DFPC refractometer with conventional (static) detection. This system could demonstrate a ($1\sigma$) sub-ppm precision for pressure assessment in the 5 kPa range.[28]

On short time scales, the assessment of refractivity from a given density of gas by the GAMOR instrumentation realized was limited by white noise. Although, in general, such noise can be averaged down, averaging can only be performed when the gas pressures have become fully equilibrated. Since it takes considerable time for the measurement cavity to reach equilibrium during the emptying stage, the performance of the original GAMOR realization on short time scales was limited by the gas evacuation process.

On long time scales, it was restricted by temperature gradients in the cavity spacer. The differences in temperature between the temperature probes and the walls of the cavities resulted in uncertainties in the assessments of the gas temperature to which the pressure is proportional.

This indicates that there are possibilities to improve on the precision of the GAMOR methodology. Regarding its short-term characteristics, it would be beneficial if the time it takes for the system to reach stable conditions for the reference assessments could be shortened, so that the reference signal could be averaged over a longer time. For the long-term performance, the precision could be improved if the signal could be averaged over longer time before being influenced by any drift.

We present in this work a novel realization of GAMOR, in part based on its original realization,[28] with some parts upgraded, but also encompassing a new gas modulation methodology, that together address these concepts. To improve on the short-term characteristics, we have developed and realized a new modulation methodology termed Gas Equilibration GAMOR (GEq-GAMOR) that shortens the time it takes to reach equilibrium conditions for the reference assessments. In this methodology, instead of evacuating the measurement cavity for the reference measurement, the gas in the measurement cavity is let into the reference cavity, to equalize the pressure in the two cavities. This implies that in GEq-GAMOR the amount of gas in both cavities are modulated within each measurement cycle. Since gas equilibration at finite pressures (above the molecular regime) is a faster process than evacuation of a cavity filled with gas down to vacuum, this allows for assessment (averaging) of the beat frequency during the reference assessment over a longer time. This reduces the noise picked up in an individual measurement cycle, which thereby contributes to an improved white noise characteristics of the system. To improve on the long-term stability, we have implemented a temperature stabilized cavity spacer with an improved temperature measurement capability.

The performance of the novel system is compared to that of the first realization of GAMOR,[28] henceforth referred to as Single Cavity Modulated GAMOR (SCM-GAMOR). It is shown that GEq-GAMOR can provide a two times better precision under short-term conditions and that the drifts have been reduced significantly (more than 8 times for integration times for 500 measurement cycles, or 18 hours). When used for low pressures, it is demonstrated that it can provide a ($2\sigma$) precision in the sub-mPa region; for an evacuated measurement cavity, and for up to *ca.* 40 measurement cycles (*ca.* 1.5 hours), the system exhibits a white-noise of 0.7 mPa (cycle)$^{1/2}$, and a minimum Allan deviation of 0.15 mPa. Finally, the system is characterized with respect to its linearity with respect to a dead weight piston gauge over the 2.8 – 10.1 kPa range.

## II. THEORY

Since GEq-GAMOR has much in common with SCM-GAMOR, the theoretical description of the former can most conveniently be based on that of the latter, which, in turn, can be assessed from expressions for ordinary (i.e. unmodulated) DFPC-based refractometry. To provide a comprehensible description of GEq-GAMOR, the basis for SCM-GAMOR is therefore reviewed in the Supplementary material [URL] ("Derivation of expressions that relate the change in beat frequency between the two laser fields to the refractivity of the gas in the measurement cavity in SCM-GAMOR and GEq- GAMOR"). Based on this, the defining expressions for GEq-GAMOR are derived and presented. In addition, the last part of the Supplementary material [URL] contains a list of the entities used together with their designation ("Nomenclature"). For both methodologies, it is assumed that, for each measurement cycle, the measurement cavity is filled with gas from an external source to a pressure of $P_{Ext}$, whose refractivity, $n_{Ext} - 1$, is to be assessed by refractometry.

## A.  SCM-GAMOR

An expression that relates the change in beat frequency between the two laser fields to the change in refractivity of the gas in the measurement cavity in SCM-GAMOR is derived as Eq. (SM-18) in the Supplementary material [URL]. This expressions shows that, when the measurement cavity, which, after being evacuated so it contains solely a minor residual amount of gas with a refractivity of $n_{Res} - 1$, is filled with gas with a refractivity of $n_{Ext} - 1$, the refractivity can be assessed from measurements of the beat frequency between the two laser fields using the expression[28]

$$n_{Ext} - 1 = \frac{\overline{\Delta f}_{(0,Res \to Ext)} + \overline{\Delta q_m}^{(Res \to Ext)}}{1 - \overline{\Delta f}_{(0,Res \to Ext)} + \varepsilon_m} + (n_{Res} - 1), \qquad (1)$$

where $\overline{\Delta f}_{(0,Res \to Ext)} = \Delta f_{(0,Res \to Ext)} / \nu_m^{(0)}$, where, in turn, $\Delta f_{(0,Res \to Ext)}$ is the shift in beat frequency when the measurement cavity, after being evacuated to a residual refractivity, is filled with the gas to be characterized. It is formally given by $f_{(0,Ext)} - f_{(0,Res)}$ where $f_{(0,Ext)}$, referred to as the filled measurement cavity beat frequency, is the beat frequency between the modes addressed in the reference and the measurement cavities when the reference cavity is empty while the measurement cavity contains the gas to be assessed (with a refractivity of $n_{Ext} - 1$), while $f_{(0,Res)}$, referred to as the evacuated measurement cavity beat frequency, is the beat frequency when the reference cavity is empty while the measurement cavity is evacuated to a refractivity of $n_{Res} - 1$. $\nu_m^{(0)}$ is the frequency of the $q_m^{(0)}$:th cavity mode addressed in the empty measurement cavity. The assessments of $f_{(0,Ext)}$ and $f_{(0,Res)}$ are henceforth referred to as the filled measurement cavity assessment and the (evacuated measurement cavity) reference assessment, respectively. $\overline{\Delta q}_m^{(Res \to Ext)}$ represents $\Delta q_m^{(Res \to Ext)} / q_m^{(0)}$, where $\Delta q_m^{(Res \to Ext)}$ is the number of modes the laser jumps when the refractivity of the measurement cavity is changed from the residual refractivity, $n_{Res} - 1$, to $n_{Ext} - 1$. $\varepsilon_m$ is the refractivity-normalized deformation coefficient of the measurement cavity,[15] defined by $\varepsilon_m (n_{Ext} - n_{Res}) = \delta L_m^{(Res \to Ext)} / L_m^{(0)}$, where $\delta L_m^{(Res \to Ext)}$ is the change in length of the cavity when it is filled with gas (under the same conditions), while $L_m^{(0)}$ is its length when being empty.

Equation (1) shows that to assess $n_{Ext} - 1$ from an measurement of $\overline{\Delta f}_{(0, Res \to Ext)}$ requires knowledge about the residual refractivity, $n_{Res} - 1$. Since the residual pressure is significantly smaller than $P_{Ext}$, it can be assessed with good accuracy by a pressure gauge. From such an assessment, $n_{Res} - 1$ can be calculated, which, in turn, can be used to recalculate the evacuated measurement cavity beat frequency, $f_{(0,Res)}$, to the empty measurement cavity beat frequency $f_{(0,0)}$. As is shown by Eq. (SM-19) in the Supplementary material [URL], by this, Eq. (1) can be simplified to

$$n_{Ext} - 1 = \frac{\overline{\Delta f}_{(0,0 \to Ext)} + \overline{\Delta q}_m^{(Res \to Ext)}}{1 - \overline{\Delta f}_{(0,0 \to Ext)} + \varepsilon_m}, \tag{2}$$

where $\overline{\Delta f}_{(0,0 \to Ext)}$ is given by $\Delta f_{(0,0 \to Ext)} / v_m^{(0)}$, where, in turn, $\Delta f_{(0,0 \to Ext)}$ is given by $f_{(0,Ext)} - f_{(0,0)}$.

Preferably, the two beat frequencies, $f_{(0,Ext)}$ and $f_{(0,0)}$ (in practice $f_{(0,Res)}$), should be assessed at the same time. However, since this is not technically possible, $f_{(0,0)}$ is, in practice, for ordinary (unmodulated) DFPC-based refractometry, measured at some instance either before or after the filled measurement cavity assessment. This introduces drifts into the system.

As has been described in the literature,[28] and as is shortly summarized in the methodology section below, the soul of the GAMOR methodology consists of a methodology to obtain an adequate estimate of the evacuated measurement cavity beat frequency at the time of the filled measurement cavity assessment. The evacuated measurement cavity beat frequency used for calculation of the shift in beat frequency is estimated by a linear interpolation from two evacuated measurement cavity reference measurements performed just prior to, and directly after, the filled measurement cavity assessment. This entity, which is denoted $\tilde{f}_{(0,0)}$ and referred to as the interpolated empty measurement cavity beat frequency, is given by Eq. (SM-11) in the supplementary material [URL] or Eq. (8) below. $\tilde{f}_{(0,0)}$ thus represents what the empty measurement cavity beat frequency would have been at the time of the filled measurement cavity assessment in case the measurement cavity had not been filled with gas. This implies that in GAMOR, $n_{Ext} - 1$ is given by Eq. (2) with $\overline{\Delta f}_{(0,0 \to Ext)}$ assessed as $\widetilde{\Delta f}_{(0,0 \to Ext)} / \nu_m^{(0)}$, where

$$\widetilde{\Delta f}_{(0,0 \to Ext)} = f_{(0,Ext)} - \tilde{f}_{(0,0)}. \tag{3}$$

## B. GEq-GAMOR

In GEq-GAMOR the filled measurement cavity assessment is performed as in SCM-GAMOR, i.e. with the measurement cavity being filled with gas while the reference cavity is evacuated. For the reference assessment, however, the gas in the measurement cavity is let into the reference cavity to equalize the pressure in the two cavities. The measured change in beat frequency in GEq-GAMOR, $\Delta f_{(Eq \to Res, Eq \to Ext)}$, therefore formally represents the difference in beat frequency between the filled measurement cavity assessment, $f_{(Res, Ext)}$, and the case when both cavities contain gas with the same but a finite refractivity, $n_{Eq}$, denoted $f_{(Eq, Eq)}$ and henceforth referred to as the gas equilibrium beat frequency, i.e. $f_{(Res, Ext)} - f_{(Eq, Eq)}$. Since $\Delta f_{(Eq \to Res, Eq \to Ext)}$ differs from the corresponding entity in SCM-GAMOR, i.e. $\Delta f_{(0, Res \to Ext)}$ or $\Delta f_{(0, 0 \to Ext)}$, neither Eq. (1), nor Eq. (2), can be used for GEq-GAMOR straight off.

There are a few possible remedies to this. The one presented here is to first recalculate $f_{(Res, Ext)}$ to $f_{(0, Ext)}$ by the use of the assessment of the residual pressure in the reference cavity (assessed by a pressure gauge), followed by a recalculation of $f_{(Eq, Eq)}$ to the empty measurement cavity beat frequency it would correspond to if both cavities would have been evacuated, denoted $f_{(0,0)}^{(Eq)}$ and referred to as the estimated empty measurement cavity beat frequency.

As is shown in the Supplementary material [URL], the latter entity is given by

$$f_{(0,0)}^{(Eq)} \approx \left[1 + \left(n_{Eq} - 1\right)\right] f_{(Eq, Eq)} - Q_{Eq} v_m^{(0)} - \Delta\varepsilon (n_{Eq} - 1) v_m^{(0)}, \quad (4)$$

where

$$Q_{Eq} = \overline{\Delta q_m}^{(0 \to Ext)} - \overline{\Delta q_m}^{(Eq \to Ext)} - \overline{\Delta q_r}^{(0 \to Eq)} v_r^{(0)} / v_m^{(0)}, \quad (5)$$

and where we have expressed the relative deformations of the two cavities due to the gas under gas equilibrium conditions, $\overline{\delta L_r}^{(0 \to Eq)}$ and $\overline{\delta L_m}^{(0 \to Eq)}$ (representing the relative deformation of the length of the reference and the measurement cavities when they are filled with gas to a refractivity of $n_{Eq}-1$, respectively), as $\varepsilon_r(n_{Eq}-1)$ and $\varepsilon_m(n_{Eq}-1)$, where, in turn, $\varepsilon_r$ is the refractivity-normalized deformation coefficient of the reference cavity, and where we have used $\Delta\varepsilon$ as a short hand notation for $\varepsilon_m - \varepsilon_r$.

This implies that, for the case with GEq-GAMOR, the refractivity of the gas that have been let into the measurement cavity can be assessed from Eq. (2) in which $\overline{\Delta f}_{(0,0\to Ext)}$ is replaced by $[f_{(0,Ext)} - \tilde{f}_{(0,0)}^{(Eq)}]/\nu_m^{(0)}$ for the elimination of drifts. $f_{(0,Ext)}$ is the residual-gas-corrected value of $f_{(Res,Ext)}$ while $\tilde{f}_{(0,0)}^{(Eq)}$ is estimated by linear interpolation {again according to Eq. (SM-11) in the supplementary material [URL] or Eq. (8) below} from two $f_{(0,0)}^{(Eq)}$ values, which, in turn, are assessed, by the use of Eq. (4), from two gas equilibrium reference assessments, i.e. two $f_{(Eq,Eq)}$, measured just prior to and directly after the filled measurement cavity assessment.

## C. Assessment of gas density and pressure

The density of the gas that have been let into the measurement cavity, $\rho_{Ext}$, can then be assessed from the refractivity by

$$\rho_{Ext} = \frac{2}{3A_R}(n_{Ext}-1)\left[1+\tilde{B}_\rho(n_{Ext}-1)\right], \tag{6}$$

where $A_R$ and $\tilde{B}_\rho$ are the molecular polarizability and the first density virial coefficient, respectively. The latter is given by $-(1+4B_R/A_R^2)/6$, where, in turn, $B_R$ is the first refractive virial coefficient in the Lorentz-Lorenz equation.[15, 30, 31] The corresponding pressure, $P_{Ext}$, can thereafter be obtained from the density as

$$P_{Ext} = k_B T N_A \rho_{Ext}[1+B_p(T)\rho_{Ext}], \tag{7}$$

where $T$ is the temperature of the gas, $N_A$ is the Avogadro's number, $k_B$ is the Boltzmann constant, and $B_p(T)$ is the first pressure virial coefficient.[26]

## III. Experimental setup

The setup for GEq-GAMOR, which is illustrated in Fig. 1, is based to a large degree on that for SCM-GAMOR.[28] Each of the two arms of the refractometer is addressed by light from a narrow linewidth Er-doped fiber laser (EDFL, NTK, Koheras Adjustik E15), emitting light within the C34 communication channel, i.e. around 1.55 µm.

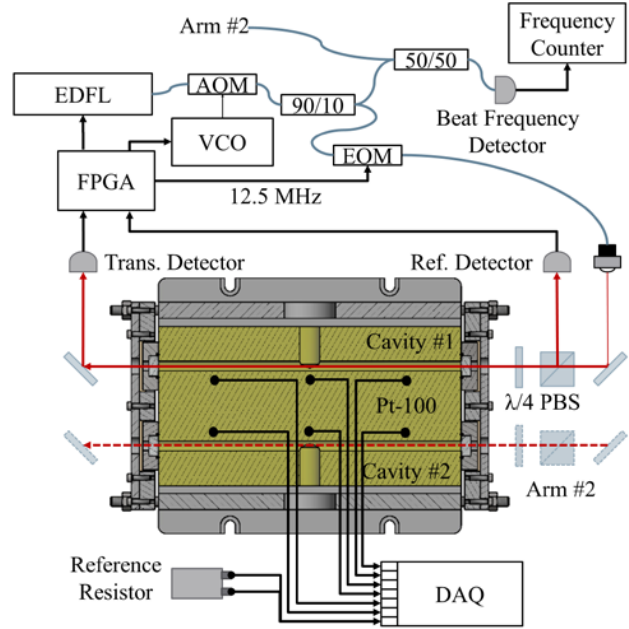

FIG. 1. Schematic illustration of the GEq-GAMOR system. Blue curves: optical fibers; red lines: free-space light propagation; black curves: wires; EDFL: Er-doped fiber laser; AOM: acousto-optic modulator; VCO: voltage controlled oscillator; 90/10: 90/10 polarization maintaining fiber splitter; 50/50: 50/50 polarization maintaining fiber splitter; Beat Frequency Detector: beat frequency detecting photo detector; EOM: electro-optic modulator; PBS: polarizing beam splitter cube; λ/4: quarter-wave plate; Ref. Detector: reflection detector; Trans. Detector: transmission detector; FPGA: field

programmable gate array; Frequency Counter: frequency counter; DAQ: data acquisition system; and Reference Resistor: standard reference resistor. Only the electronics and some parts of the optics of one arm (#1) is shown. See Silander et al. for details of the omitted parts.[28] The left out parts for arm 2 are identical to those of arm 1.

For improved locking bandwidth, the light from each laser is sent through a fiber-coupled acousto-optic modulator (AOM, AA Opto-Electronic, MT110-IR25-3FIO). The first order output from this is split by a 90/10 polarization maintaining fiber splitter (Thorlabs, PMC1550- 90B-FC), whose high transmission output is connected to an electro-optic modulator (EOM, General Photonics, LPM- 001-15) that is modulated at 12.5 MHz for Pound-Drever-Hall (PDH) locking. The output of the EOM is then coupled into free space by a collimator and sent through a mode-matching lens, a polarizing beam splitter (PBS), and a quarter-wave ($\lambda/4$) plate, before it enters the cavity. The low power outputs of the 90/10 fiber splitters of the two arms are merged by a 50/50 fiber coupler (Thorlabs, PMC1550-50B-FC) whose output is sent to a beat frequency detecting photo detector (Thorlabs, PDA8GS).

For the locking, both the back reflected and the transmitted light are used. The reflected light is passed through the quarter-wave plate a second time and is deflected by the PBS before it is collected with a fast photo detector (Thorlabs, PDA10CE-EC). The transmitted light is detected by a large area photo detector (Thorlabs, PDA50B-EC).

The outputs from the photo detectors are connected to a commercial digital servo module based on a field programmable gate array (FPGA, Toptica, Digilock 110). In this, to produce a PDH error signal for the locking, the reflected signal is demodulated at 12.5 MHz before it passes through one of two proportional–integral–derivative (PID) servos. One provides a slow feedback (with a bandwidth up to 100 Hz), which controls the piezoelectric transducer of the fiber laser, while the other gives a fast feedback (with a bandwidth up to around 100 kHz), which is connected to a voltage controlled oscillator (VCO) that produces an RF frequency of around 110 MHz for the frequency tuning of the first order output of the AOM. The transmission signal is used to activate feedback to the laser to enable automatic re-locking during controlled mode jumps.

The output of the FPGA that is routed to the laser is limited to a voltage corresponding to a change of the laser frequency of two FSR. When the feedback voltage reaches this limit, the automatic re-locking routine of the module relocks the laser to an adjacent mode with a frequency closer to the center of its working range. The relocking process is fast, it takes typically a tenth of a second, and it does not influence the refractivity assessment (since no mode jumps take place during data acquisition). Hence, it allows for a dynamic range that is not limited by the tunability of the lasers.

The output signal of the beat frequency detecting photo detector is sent to a frequency counter (Freq. Counter, Aim-TTi instruments, A TF960). The beat frequency is acquired with a rate of 5 Hz. As the frequency counter is limited to 6 GHz, temperature tuning is used to initially (i.e. before the measurement series) set the frequencies of the two lasers so that the beat frequency is in the center of the range of the frequency counter. After this manual (coarse) setting, the automatic re-locking routine keeps the beat frequency between 2 and 6 GHz under all measurement conditions.

As an upgrade compared to the previous GAMOR system,[28] the temperature of the cavity spacer is monitored by six Pt-100 sensors (RS Pro PT100 Sensor, 457-3710) that are placed in holes drilled into the cavity spacer for monitoring of the temperature and its distribution in the spacer. The holes were bored so that the distance between the sensors and the measurement cavity wall is 5 mm. The sensors are connected to two DAQ (data acquisition) modules (National instruments universal analog input module, NI-9219). The probes and the DAQ modules were calibrated at RISE to within a combined uncertainty of 10 mK. During operation, both the (temporal) stability and the (spatial) gradients in temperature of the cavity were assessed to be below 5 mK. To monitor drifts in the DAQ modules a 100 Ω standard reference resistor is simultaneously monitored.

The DFPC is connected to a gas and vacuum system whose main parts are displayed in Fig. 2. The system comprises three parts: a gas supply unit (the leftmost part of the system depicted in Fig. 2), a pressure stabilizing unit, providing a stable pressure (the center part of the system), and the refractometer (the rightmost part of the system).

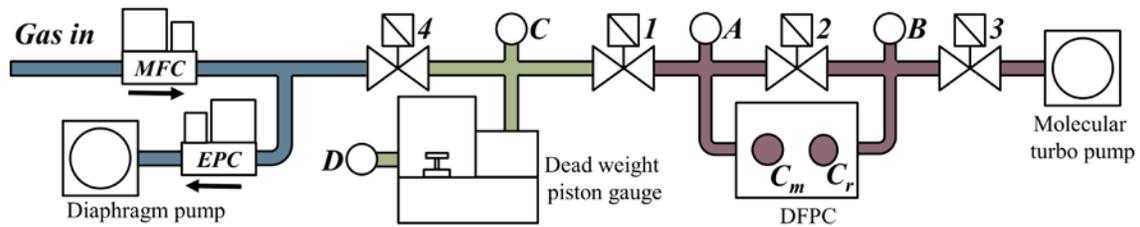

FIG. 2. Schematic illustration of the vacuum system in GEq-GAMOR. The system comprises three parts: a gas supply unit (the leftmost part of the system, blue gas lines), a pressure stabilizing unit (the center part of the system, green gas lines), and the refractometer (the rightmost part of the system, red gas lines). MFC and EPC are a mass flow controller and an electronic pressure controller, respectively. The pressure is regulated by a dead weight piston gauge. 1, 2, 3, and 4 represent solenoid valves. A, B, C, and D indicate pressure gauges. $C_m$ and $C_r$ illustrate the measurement and reference cavities of the DFPC, respectively. The system is evacuated through valve 3 by a turbo vacuum pump. (*To be printed with double column width*)

High-purity nitrogen gas is supplied from a central gas unit (represented by "Gas in" in the figure) into the gas supply unit. To ensure stable conditions, a mass flow controller (MFC, Bronkhorst, F-201CV) and an electronic pressure controller (EPC Bronkhorst, P-702CV) are used to provide a continuous flow of gas through the gas supply unit (to reduce the risk for gas contamination) at a constant pressure.

When the pressure assessed by pressure gauge C (Oerlikon-Leybold CTR 101N 1000 Torr) is below a set threshold pressure, valve 4 (Oerlikon-Leybold 28444) opens to fill the pressure stabilizing unit, consisting of a dead weight piston gauge (RUSKA 2465A), with gas. When the pressure reaches a second threshold, set to the piston gauge pressure ($P_{Ext}$), valve 4 closes, whereby the piston gauge floats, providing a stabilized pressure. Pressure gauge D (Oerlikon-Leybold CTR 101N 0.1 Torr) is used to monitor the hood pressure of the dead weight piston gauge.

The refractometer comprises the DFPC, three solenoid valves (1 and 2: Oerlikon-Leybold 28444; and 3: Leycon 215006v01), which are used to control the flow of gas in and out of the cavities, and two pressure gauges (A: Oerlikon-Leybold CTR 101N 1000 Torr; and B: Oerlikon-Leybold CTR 101N 0.1 Torr), which are used to monitor the pressure in the two cavities. During certain parts of the measurement cycle (see below), the gas system is evacuated through valve 3 by a turbo vacuum pump.

## IV. Methodology

Before any measurement series was initiated, the wavelengths of the lasers addressing the two cavities under evacuated conditions were measured with a wavelength meter (Burleigh, WA-1500) and converted to frequency, $v_m^{(0)}$ and $v_r^{(0)}$, respectively. The free spectral range (FSR) of each cavity was measured as the frequency shift of the corresponding laser when it was exposed to controlled cavity mode jumps. These two assessments provide sufficient information to unambiguously determine the empty cavity mode numbers $q_m^{(0)}$ and $q_r^{(0)}$ that were addressed by the lasers under evacuated conditions.

To assist in the assessment of mode jumps (see below), the lasers were characterized with respect to their frequency-to-voltage responses, i.e. to $v_m = f(V_m)$ and $v_r = f(V_r)$, where $V_m$ and $V_r$ are the voltages sent to the PZTs of the two lasers.

## A. The valve switching and gas modulation procedure

As for SCM-GAMOR, to achieve gas modulation in the system, GEq-GAMOR is realized by periodically changing the states of the various valves in the system. Figure. 3 shows a schematic illustration of the valve states for the three different states that GEq-GAMOR comprises.

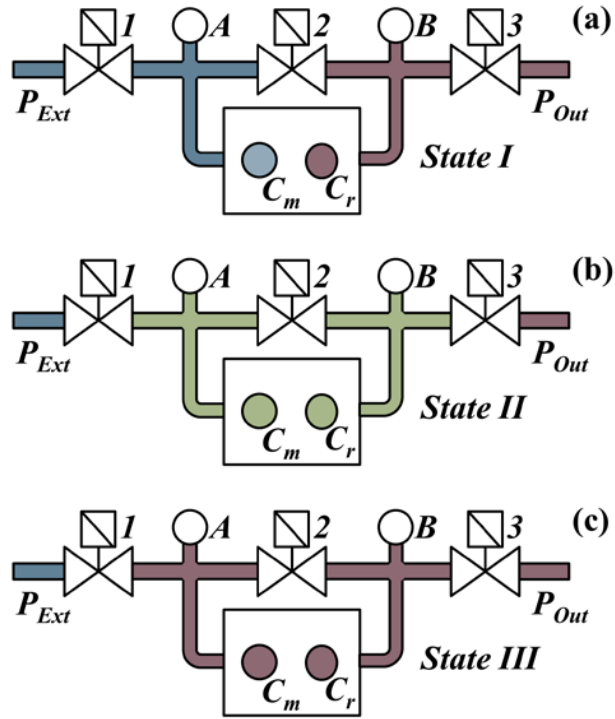

FIG. 3. Schematic illustration of the various states of the GEq-GAMOR. Panel (a): State I starts with valve 2 being closed, after which valve 1 is opened. The pressure in the measurement cavity, $C_m$, increases until it becomes equal to the pressure of the dead weight piston gauge, $P_{Ext}$. The pressure in the reference cavity $C_r$, is equal to the pressure on the output port $P_{Out}$, which, during pumping, represents the residual pressure in $C_r$. Panel (b): State II is initiated by closing the valves 1 and 3, whereafter valve 2 is opened so that the pressures in $C_m$ and $C_r$ equilibrate. Panel (c): State III begins with valve 3 being opened to evacuate both cavities. The gas lines with the same color have the same pressure. Hence, a valve that has the same color of the gas lines on its two sides is open, while, if the colors are dissimilar, it is closed.

In *State I*, the beat frequency for the filled measurement cavity assessment, i.e. $f_{(0,Ext)}$, is assessed. State I follows state III of the previous cycle (see below), in which both cavities are evacuated (obtained by having valves 2 and 3 open). As is illustrated in Fig. 3(a), state I is initiated by closing valve 2 and opening valve 1. As is shown in section I of Fig. 4(a), by this, the measurement cavity ($C_m$) is filled with gas up to the pressure that is supplied by the dead weight piston gauge, which is set to $P_{Ext}$, while the reference cavity ($C_r$) is being evacuated. To assess the residual refractivity in the reference cavity, $(n_{Res} - 1)$, pressure gauge B is used to measure the residual pressure in the cavity.

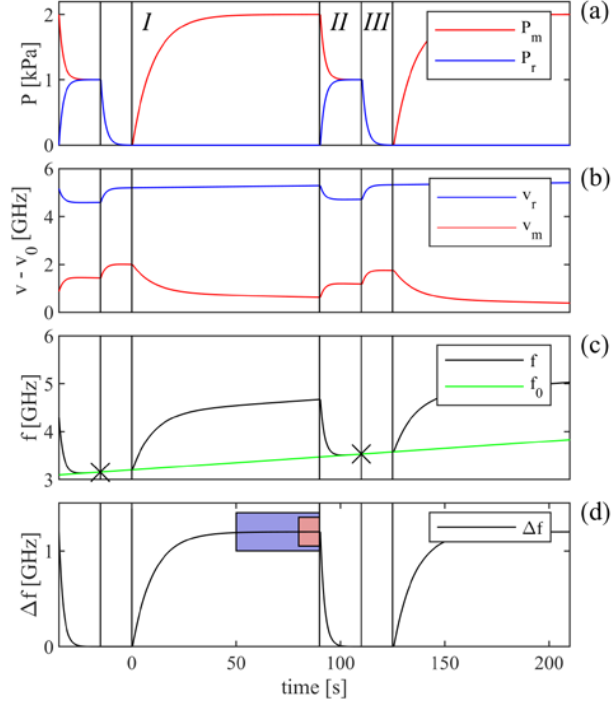

FIG. 4. The principle of the GEq-GAMOR methodology displayed over two full measurement cycles. Panel (a) displays, as functions of time, the pressures in the measurement cavity, $P_m(t)$, (red curve) and in the reference cavity, $P_r(t)$, (blue curve). Panel (b) shows the corresponding frequencies of the measurement and reference lasers, $v_m(t)$ (red curve) and $v_r(t)$ (blue curve), respectively, in the absence of mode jumps (for display purposes, both offset to a common frequency $v_0$). Panel (c) illustrates the corresponding beat frequencies: the measured one, $f(t)$ (black curve), in the legend for simplicity denoted $f$, and the estimated empty measurement cavity beat frequency, $f_{(0,0)}^{(Eq)}(t)$ (green line), in the legend denoted $f_0$, whose slope is greatly exaggerated for display purposes. Panel (d), finally, displays the cavity-drift-corrected shift in beat frequency, $\Delta f(t)$, (black curve), given by the difference between $f(t)$ and the interpolated $f_{(0,0)}^{(Eq)}(t)$. The large colored box in panel (d) indicates over which interval the filled measurement cavity beat frequency is evaluated while the smaller box shows the time period over which the data alternatively were evaluated to assess the dependence of the averaging time on the noise (see the text for details).

While in state I, the pressure in the measurement cavity is measured with pressure gauge A. Based on this, together with the voltage sent to the PZT of the laser addressing the measurement cavity when the pressure has settled, $V_m$, the number of mode jumps performed when the gas was let into the measurement cavity, $\Delta q_m^{(Res \to Ext)}$, can be unambiguously assessed.

In *State II,* as is shown in Fig. 3(b), the valves 1 and 3 are closed whereafter valve 2 is opened. As is illustrated in section II in Fig. 4(a), by this, the pressures in the two cavities are equilibrated. This provides conditions for assessment of the gas equilibrium beat frequency, $f_{(Eq,Eq)}$. Simultaneously, pressure gauge A is used to measure the pressure from which an approximate value for $(n_{Eq} - 1)$ [to be used in Eq. (4)] can be estimated. [Since the $(n_{Eq} - 1)$-terms in Eq. (4) typically only contribute to the value of $f_{(0,0)}^{(Eq)}$ on the 10$^{-5}$ level (on a relative scale), it suffices to assess $(n_{Eq} - 1)$ with two significant digits to assess $f_{(0,0)}^{(Eq)}$ with sub-ppm accuracy.] As in state I, the pressure readings in combination with $V_m$ and $V_r$ are used to determine $\Delta q_m^{(0 \to Eq)}$ and $\Delta q_r^{(0 \to Eq)}$. All this provides conditions for assessment of $f_{(0,0)}^{(Eq)}$ by the use of Eqs. (4) and (5) with good accuracy.

This also indicates that there is no need to continuously or actively monitor the number of mode jumps during any gas filling or evacuation process; the status of the system provides, at any time, enough information to deduce any mode jump in any cavity.

In *State III*, to ensure gas purity (i.e. to reduce the influence from outgassing and gas leaks), as is shown in Fig. 3(c), the state begins by opening valve 3 whereby, as is shown in section III in Fig. 4(a), both cavities are evacuated.

After this, state I of the next cycle starts.

For this particular case, when the pressure to be assessed is provided by a dead weight piston gauge (as is shown by Fig. 2), the filling of the piston gauge (as was described above, controlled by valve 4), takes place during the first part of state I.

Figure 4(b) shows the frequencies of the measurement and reference lasers, $v_m(t)$ (red curve) and $v_r(t)$ (blue curve), respectively, for simplicity, in the absence of mode jumps, during the various states (for display purposes, both are offset to a common frequency $v_0$).

## *B. The GAMOR feature*

The beat signal between the two laser fields is in practice measured during the entire measurement cycle. It can therefore be seen as a continuous function of time, i.e. as $f(t)$. A schematic representation of a possible $f(t)$ cycle is given by the black curve in Fig. 4(c). As has been alluded to above, because of drifts of the cavity spacer, this signal will not be a complete replica of the gas pressure [schematically displayed by the red curve in Fig. 4(a)]. To alleviate this, the estimated empty measurement cavity beat frequency, $f_{(0,0)}^{(Eq)}(t)$, is calculated at all instances of the measurement cycle by use of a linear interpolation between two estimated empty measurement cavity frequencies. The first is measured before the filling of cycle *n*, $f_{(0,0)}^{(Eq)}(t_n)$, and the second after the cycle [identical to the one before the filling of cycle *n+1*], $f_{(0,0)}^{(Eq)}(t_{n+1})$, marked by crosses in Fig. 4(c). In practice, in GEq-GAMOR, these are obtained by assessment of two gas equilibrium reference assessments, $f_{(Eq,Eq)}$, at the two time instances $t_n$ and $t_{n+1}$, i.e. $f_{(Eq,Eq)}(t_n)$ and $f_{(Eq,Eq)}(t_{n+1})$, respectively, and the use of Eq. (4). This implies that the interpolated estimated empty measurement cavity beat frequency, denoted $\tilde{f}_{(0,0)}^{Eq}(t)$ and marked by the green curve in in Fig. 4(c), can be assessed, at any time within each cycle (for cycle $n$, for which $t_n \leq t \leq t_{n+1}$), as

$$\tilde{f}_{(0,0)}^{(Eq)}(t) = f_{(0,0)}^{(Eq)}(t_n) + \frac{f_{(0,0)}^{(Eq)}(t_{n+1}) - f_{(0,0)}^{(Eq)}(t_n)}{t_{n+1} - t_n}(t - t_n). \tag{8}$$

A cavity-drift-corrected shift in the beat frequency, corresponding to the gas in the cavity, $\Delta f(t)$, is then calculated as the difference between the measured beat frequency, $f(t)$ (black curve), and the interpolated estimated empty measurement cavity beat frequency, $\tilde{f}_{(0,0)}^{(Eq)}(t)$ (green straight line), both in Fig. 4(c). The cavity-drift-corrected shift in beat frequency, which is represented by the black curve in Fig. 4(d), then constitutes (after correction for possible mode jumps), by the use of Eq. (2), a representation of the refractivity in the measurement cavity, $(n_m - 1)(t)$, during the entire cycle. By use of the Eqs. (6) and (7), this provides also information about the momentary density, $\rho_m(t)$, and pressure, $P_m(t)$, at all times in the measurement cavity.

The values of the molecular polarizability, $A_R$, and the density and pressure viral coefficients, $\tilde{B}_R$ and $B_p(T)$, were taken as those used in original SCM-GAMOR work,[28] which, as is described in that work, in turn, are based upon the literature,[14, 26, 32] and where the molecular polarizability additionally has been recalculated for the actual wavelength used. The two refractivity-normalized deformation coefficients, $\varepsilon_m$ and $\varepsilon_r$, have, in this work, been set to zero since the DFPC has not yet been fully characterized.

The average of the values of the cavity-drift-corrected shift in beat frequency, $\Delta f(t)$, during the part of the cycle when the cavity pressure, $P_m$, has been equalized with respect to the external pressure, $P_{Ext}$, which are marked by the large colored box at the end of section I of Fig. 4(d), represents the $\Delta f_{(0,0 \to Ext)}$ entity. This is then used to assess the refractivity according to the Eqs. (1), (4), and (5), as well as the density and pressure, $\rho_{Ext}$ and $P_{Ext}$, by additional use of the Eqs. (6) and (7).

Hence, by this procedure, the effect of the linear drift of the cavities is efficiently eliminated from the assessment of gas refractivity, density, and pressure.

## V. Results and Discussion
### A. *Stability and precision*

To evaluate the stability and precision of the refractometer it was compared to a dead weight piston gauge (RUSKA 2465A) set to a pressure of 4303 Pa ($P_{Ext}$). Figure 5 shows the pressure difference $\Delta P$ between the pressure assessed by the refractometer, $P_R$, and that set of the dead weight piston gauge, $P_{Ext}$. The data in Fig. 5(a) consist of 720 individual measurements cycles (each lasting 100 s) taken over a period of 20 h assessed with a SCM-GAMOR system (blue markers), while those in Fig. 5(b) comprise 2700 measurements cycles taken over a period of 100 h (each lasting 130 s) for the GEq-GAMOR system (red markers). The integration times for the filled measurement cavity assessments were 30 and 40 s for SCM-GAMOR and GEq-GAMOR, respectively. Since neither the dead weight piston gauge, nor the DFPC, has yet been fully characterized, the means of the difference-signals have been subtracted. The dashed horizontal lines represent two standard deviations of the pressure differences, i.e. ±2σ.

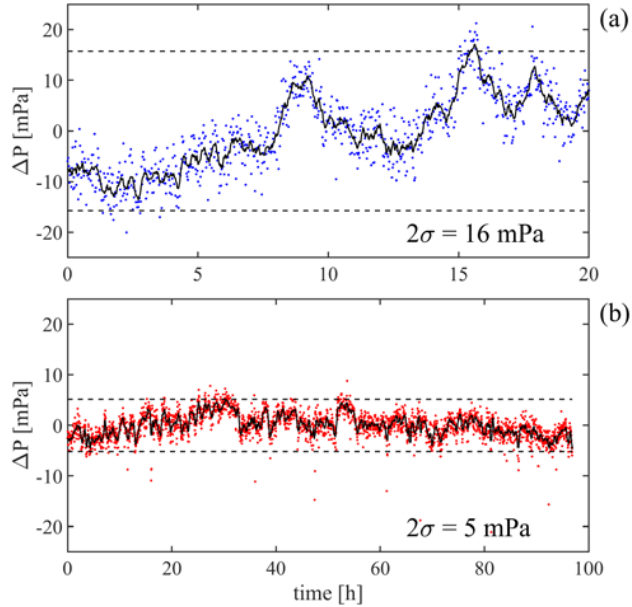

FIG. 5. Stability measurement comprising the difference in pressure, $\Delta P$, between a refractometer ($P_R$) and the dead weight piston gauge set to 4303 Pa ($P_{Ext}$). Panel (a) shows, by the blue markers, the case for the SCM-GAMOR system taken over 20 h (data adapted from Silander et al.[28]) while panel (b) displays, by the red markers, that for the GEq-GAMOR system measured over 100 h. As a help to guide the eye, the black solid curves represent the moving mean over 10 samples. The dashed horizontal lines represent two standard deviations of the pressure differences, i.e. ±2σ.

The data taken by the GEq-GAMOR system [Fig. 5(b)] show that the combined 2σ stability of the system (representing the fluctuations between the deadweight piston gauge and the refractometer) over 4 days was within ±5 mPa (or ±1 ppm). This is a significant improvement from the original realization of GAMOR [the SCM-GAMOR system, Fig. 5(a)], which showed ±16 mPa (or ±4 ppm) over 20 h (data adapted from Silander et al.[28]). This indicates that the GEq-GAMOR system has a standard deviation measured over 100 h that is more than a factor of three smaller than that of the SCM-GAMOR system assessed over 20 h. This improvement is attributed to the upgraded temperature control and assessment.

Figure 6 shows, by the blue and the red markers, i.e. the first and third sets of data counted from above, respectively, the Allan deviation of the data presented in Fig. 5 (the SCM- and the GEq-GAMOR systems, respectively). The plots show that there is a clear improvement in the stability on all time scales. For comparison, the purple markers (the lowermost curve) represent a measurement series with an evacuated measurement cavity.

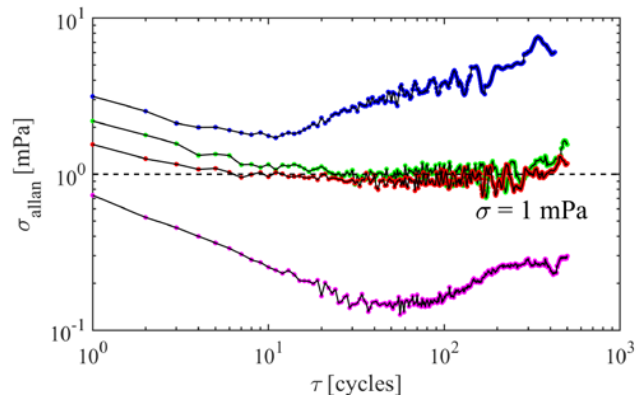

FIG. 6. Allan deviation of the two systems: blue markers (the uppermost data set): SCM-GAMOR,[28] red markers (the third set of data counted from above): GEq-GAMOR, both taken for a set pressure ($P_{Ext}$) of 4303 Pa. The green markers (the second set of data counted from above) represent the GEq-GAMOR data evaluated with the integration times of the residual gas pressure measurement and the beat frequency reduced by a factor of four. The dashed line represents an Allan deviation of 1 mPa. The purple markers (the lowermost set of data) represent zero pressure measurement.

It can be seen by the leftmost data points, which represent the short-term response of the system, that the novel GEq-GAMOR system, provides, for averaging times below 10 measurement cycles, in comparison with the original SCM-GAMOR system, a reduction in noise by a factor of 2, from 3 to 1.5 mPa $(cycle)^{1/2}$. This is attributed to the longer integration time in the reference pressure assessment, which reduces white-noise-type fluctuations.

To verify this assumption, the data for the GEq-GAMOR system (the red markers), which were integrated for 40 s, were re-evaluated with a reduced integration time (10 s), corresponding to the smaller box in Fig. 4(d). The resulting set of data is presented by the green markers in Fig. 6 (the second set of data counted from above). A comparison between these two GEq-GAMOR data sets (green and red) indisputably shows that the integration time affects the short-term response of the system. Assuming that both curves are affected by the same amount of flicker noise for the shortest time scales as for the longer, i.e. 1 mPa (see below), the residual white-noise contribution to the short-term response in the two data sets can be estimated to be 2 and 1 mPa (cycle)$^{1/2}$, respectively. This is in agreement with the expected improvement of a factor of 2 originating from the square root of the decreased integration time.

Regarding the long-term response, the data show that the GEq-GAMOR system does not exhibit the same amount of drifts for longer averaging times (above 20 cycles) as the SCM-GAMOR system does. For integration times in the 300 – 500 cycle (11 – 18 hours) interval, the Allan deviation of the GEq-GAMOR system is 6 – 8 times lower than for the SCM-GAMOR system. This reduction in drift is attributed to the improved temperature measurement and control in the GEq-GAMOR setup.

The data show though that, for a pressure of 4303 Pa, the GEq-GAMOR system is limited by flicker noise for averaging times above 10 cycles. The flicker noise, expressed in terms of an Allan deviation, is 1 mPa (which corresponds to 0.25 ppm) for averaging times up to 500 cycles (or 18 h). This corresponds to a precision, defined as twice the Allan deviation, of 0.5 ppm. Although the origin of this flicker noise has not been irrefutably identified, one possible reason is that it originates from the non-linear parts of the drifts that the GAMOR methodology does not eliminate.[33]

The purple markers, which represent a measurement series with an evacuated measurement cavity, indicate a white-noise limited Allan deviation, up to around 40 cycles, of 0.7 mPa $(cycle)^{1/2}$, and, for averaging times in the 40 – 80 cycle range (corresponding to 1.5 – 3 hours), a minimum Allan deviation of 0.15 mPa. The white-noise limited deviation is a factor of 2 better than what was achieved with the SCM-GAMOR system [which was assessed to 1.4 mPa $(cycle)^{1/2}$].[28] This agrees well with the corresponding difference in averaging time of the beat frequency for the reference assessments for SCM-GAMOR and GEq-GAMOR (0.2 s for the evacuated measurement cavity assessment for SCM-GAMOR and 0.8 s for the equilibration beat frequency assessment for GEq-GAMOR, respectively). This shows that the white noise during the reference measurement is in the order of 0.6 mPa $Hz^{-1/2}$ for both methodologies.

However, when assessments are performed with finite (non-zero) pressures, the white noise is higher than when the measurement cavity is empty. For example, for a pressure of 4303 Pa, the white noise was found to be around 6 mPa $Hz^{-1/2}$. The additional noise is believed to originate from a combination of fluctuations in pressure from the dead weight piston gauge and temperature (including its assessment).

## *B. Linearity*

To assess the linearity of the system a series of measurements were performed for a set of piston gauge weights. Figure 7(a) presents, by the individual markers, the external pressure assessed by use of the refractometer, denoted $P_R$, as a function of the estimated pressure of the piston gauge, $P_{dw}$, calculated by use of Eq. (4) in Silander et al.[28], for seven different weights. The data series were taken over a period of 8 days in a non-consecutive order (2841, 4303, 7225, 10148, 3426, 5764, and 8687 Pa). Fig. 7(b) displays the averages of the same set of data in relative units.

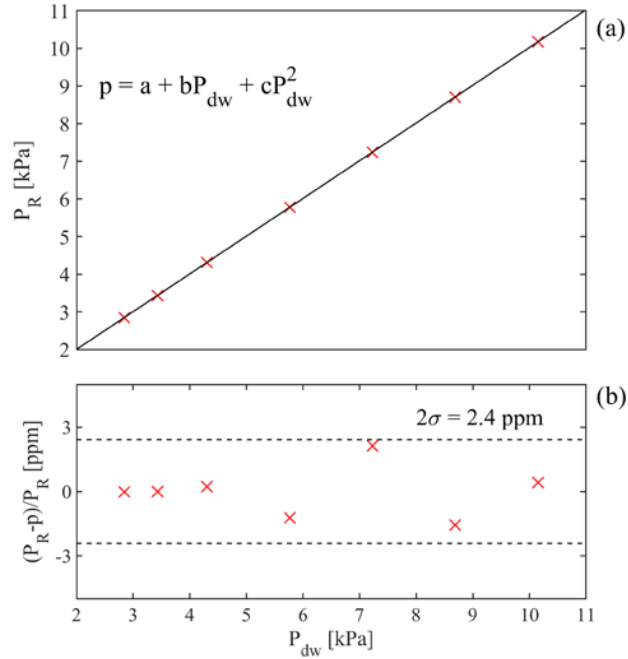

FIG. 7. Characterization of the response of the GEq-GAMOR refractometer, $P_R$, with respect to a dead weight piston gauge, $P_{dw}$. Panel (a): The pressure assessed with the refractometer, $P_R$, as a function of the estimated pressure of the dead weight piston gauge, $P_{dw}$, for seven different weights. Each red marker represents a set of 50 measurement cycles that have a spread that are significantly smaller than the size of the markers. Hence, they appear in panel (a) as a single marker. The black curve represents a second order or a linear fit to the data (see text for details), denoted $p$. Panel (b): The deviation of the average values of the various sets of measurement cycles from the pressures estimated by the linear fit displayed on a relative scale. The dashed horizontal lines represent two standard deviations (i.e. $\pm 2\sigma$) of the entire set of data (2 ppm).

Regarding the data in Fig. 7(a), the black curve shows the best second order fit to the data of the type

$$p = a + bP_{dw} + cP_{dw}^2. \tag{9}$$

The fitted function agrees very well with the data. It addition, it provides a value of the $c$ coefficient of $3 \times 10^{-10}$ Pa$^{-1}$. This shows that the GEq-GAMOR system provides a response that has a very small (virtually insignificant) non-linear component; the value of the non-linear term is, for each pressure, smaller than the estimated noise level. It can therefore be concluded that the GEq-GAMOR system, evaluated by the theory and the evaluation procedures described above, does not exhibit any noticeable systematic non-linear dependence over the pressure range addressed. Hence, it suffices to evaluate its performance with respect to its linear response.

A corresponding linear fit to the data in Fig. 7(a), visually indistinguishable from the fit displayed in the same panel, shows that the coefficients $a$ and $b$ are 0.108(6) Pa and 1.002280(1), respectively, where the uncertainties are estimated from the spread in data at each pressure.

This shows that the deviation of the $b$ coefficient from unity ($2.280 \times 10^{-3}$) is systematic. Possible contribution to this discrepancy is presumed to be the omission of the cavity deformation (represented by the $\varepsilon_r$ and $\varepsilon_m$ entities in the expressions previously addressed) in the assessment of refractivity and pressure from the data by use of the expressions above, the molar polarizability, the temperature assessment, and an incorrect characterization of the dead weight piston gauge.

The fit also reveals a non-zero value of the constant term. The cause for this is presently unknown, but possible causes can be an incorrect piston weight or the assessments of the hood and residual pressure.

## VI. Summary, Conclusions, and Outlook

A novel technique for refractometry, based on Gas Modulation Refractometry (GAMOR), that outperforms the original realization of GAMOR, here referred to as Single Cavity Modulated GAMOR (SCM-GAMOR), has been developed, realized, and scrutinized. It is based upon the fact that the reference measurements, which in SCM-GAMOR are performed by evacuating the measurement cavity, instead are carried out by equalizing the pressures in the two cavities. This new methodology is therefore referred to as Gas Equilibration GAMOR (GEq-GAMOR).

By this, the time it takes to reach adequate conditions for the reference measurements has been reduced. This implies that a larger fraction of the measurement cycle time can be devoted to acquisition of data, in particular during the reference assessments. In addition, the residual gas pressure assessment, which for GEq-GAMOR is performed during state I, can be averaged significantly longer than what is the case for SCM-GAMOR (in which the same entity had to be measured under non-equilibrium conditions in state II while the system is being pumped down).

Both these features reduce the white noise and improve on the short-term characteristics.

Yet another advantage of GEq-GAMOR is that the pressure during the reference measurement (i.e. the equilibration pressure) does not need to be assessed with the same accuracy as the residual pressure in the measurement cavity in SCM-GAMOR.

The system presented also incorporates a new cavity design with improved temperature stabilization and assessment. This has contributed to an improved long-term response of the GAMOR methodology.

A characterization of the novel GEq-GAMOR system, by use of a dead weight piston gauge, shows that it can provide a short-term response that exceeds that of the original SCM-GAMOR system by a factor of two. For longer integration times (above 10 cycles, *ca.* 20 min), and for a pressure of 4303 Pa, the system is limited by flicker noise, which, when expressed in terms of an Allan deviation, is 1 mPa, which, in turn, for this pressure, corresponds to 0.25 ppm. This thus corresponds to a precision (defined as twice the Allan deviation) of 0.5 ppm. For the case with an evacuated measurement cavity, it was found that the system is white-noise limited up to around 40 cycles (*ca.* 1.5 hours), with a white-noise of 0.7 mPa (cycle)$^{1/2}$, and that it exhibits a minimum Allan deviation (for averaging times in the 40 – 80 cycle range, corresponding to 1.5 – 3 hours) of 0.15 mPa. For the longest integration times considered, and again for a pressure of 4303 Pa, it was found that the GEq-GAMOR system could reduce long-term drifts with respect to that of the original SCM-GAMOR system significantly; for 18 hours, by a factor of 8.

The linearity of the system was assessed by a comparison with a dead weight piston gauge. It was found that the GEq-GAMOR system provides a linear response (with respect to the piston gauge) over the entire pressure range investigated (2.8 – 10.1 kPa), with no evidence of any systematic non-linearity. This range can be extended downwards into the Pa region with an expected precision in the sub-mPa region. This implies that the system can be used for transfer of calibration for pressures outside the range in which it was characterized.

Outgassing and leaks will contaminate the gas over time and affect its refractivity.[14] This will degrade the accuracy of density and pressure measurements. As the GAMOR principle involves periodical evacuation of the cavities, as is discussed in some detail elsewhere, the effect of gas contamination in GAMOR methodologies is assumed to be small.[33] If though still non-negligible, it can be quantified by altering the modulation period and interpolating the modulation period to zero. This provides a system that in practice is not affected by leakages or outgassing.[33]

This work has mainly been focused on improving the precision and long-term stability of GAMOR. The accuracy of the instrumentation has not yet been addressed. In this case, it was found that the system has a finite (systematic) discrepancy with respect to the dead weight piston gauge, with a linear response that shows a deviation of 0.228 % from the ideal response (unity). This discrepancy can be attributed to a number of causes of which one is the un-characterized DFPC (the deformation of the cavity due to the pressure of the gas, which has not yet been assessed).

In the future, the accuracy of the system can be improved by a characterization with respect to both the physical deformation of the cavity, i.e. the value of $\varepsilon_m$ and $\Delta\varepsilon$, and the gas constants. This can, for example, be done by the use of two gases assessed under the same conditions, of which one is He, for which the molar refractivity (together with the relevant density and pressure virial coefficients) is known. Alternatively, the system can be calibrated in terms of the combination of molar refractivity and physical elongation of the cavity by the use of a pressure standard.

Since it has been shown that it is possible to construct a GAMOR setup from off-the-shelf components with a relatively simple cavity design, this opens up for a new class of pressure standards, with no moving actuators, that are traceable to the SI system through frequency and temperature. The potential benefits of these standards, in comparison to current mercury manometer and dead weight piston gauges, are substantial, both in reduced maintenances and operating cost.

# ACKNOWLEDGMENTS

This research was in part supported by the EMPIR initiative, which is co-founded by the European Union's Horizon 2020 research and innovation program and the EMPIR Participating States; the Swedish Research Council (VR), project number 621-2015-04374; Umeå University Industrial doctoral school (IDS); Vinnova Metrology Programme, project numbers 2015-0647 and 2014-06095; and the Kempe Foundations, project number 1823, U12. The authors are also indebted to Magnus Holmsten at RISE for calibrating the temperature sensors.

# Supplementary material to

## Gas equilibration gas modulation refractometry (GEq-GAMOR) for assessment of pressure with sub-ppm precision

Isak Silander, Thomas Hausmaninger, Clayton Forssén,

Department of Physics, Umeå University, SE-901 87 Umeå, Sweden

Martin Zelan [a],

Measurement Science and Technology, RISE Research Institutes of Sweden, SE-501 15 Borås, Sweden

Ove Axner [b]

Department of Physics, Umeå University, SE-901 87 Umeå, Sweden

[a] Electronic mail: martin.zelan@ri.se
[b] Electronic mail: ove.axner@umu.se

# DERIVATION OF EXPRESSIONS THAT RELATE THE CHANGE IN BEAT FREQUENCY BETWEEN THE TWO LASER FIELDS TO THE REFRACTIVITY OF THE GAS IN THE MEASUREMENT CAVITY IN SCM-GAMOR AND GEQ-GAMOR

**SCM-GAMOR**

In SCM-GAMOR (Single Cavity Modulated GAMOR), the measurement cavity is alternatingly evacuated and filled with gas whose refractivity (density or pressure) is to be assessed while the reference cavity is held at a constant pressure. This implies that the frequency of the laser that addresses the measurement cavity (henceforth referred to as the measurement laser) alternates between two frequencies, one when the measurement cavity is being evacuated (referred to as the reference assessment) and another when it contains gas (referred to as the filled measurement cavity assessment).

To relate such a change in frequency to the corresponding change in refractivity, let us first, to set the nomenclature, assume that the frequency of the laser that addresses the measurement cavity when it is fully evacuated (to a pure vacuum), $v_m^{(0)}$, can be written as

$$v_m^{(0)} = \frac{q_m^{(0)} c}{2 L_m^{(0)}}, \qquad (\text{SM-1})$$

where $q_m^{(0)}$ is the number of the mode to which the measurement laser is locked, $L_m^{(0)}$ is the length of the measurement cavity, both when the cavity is empty, and $c$ is the speed of light in vacuum.

Since, for practical reasons, the measurement cavity is not evacuated to a pure vacuum during the reference assessment, we will assumed that, under this assessment, it contains a residual amount of gas that has a refractivity of $n_{Res} - 1$. The corresponding frequency of the cavity mode will be denoted $v_m^{(Res)}$. The corresponding entities for the case when the measurement cavity is filled with the gas whose refractivity is to be assessed (referred to as the filled measurement cavity assessment) are $n_g - 1$ and $\nu_m^{(g)}$, respectively. Based on Eq. (SM-1) as well as other sources,[1] this implies that $\nu_m^{(Res)}$ and $\nu_m^{(g)}$ can be expressed as

$$\nu_m^{(Res)} = \frac{\left[q_m^{(0)} + \Delta q_m^{(0 \to Res)}\right]c}{2n_{Res}\left[L_m^{(0)} + \delta L_m^{(0 \to Res)}\right]} = \nu_m^{(0)} \frac{1 + \overline{\Delta q_m}^{(0 \to Res)}}{n_{Res}\left[1 + \overline{\delta L_m}^{(0 \to Res)}\right]} \quad \text{(SM-2)}$$

and

$$\nu_m^{(g)} = \frac{\left[q_m^{(0)} + \Delta q_m^{(0 \to g)}\right]c}{2n_g\left[L_m^{(0)} + \delta L_m^{(0 \to g)}\right]} = \nu_m^{(0)} \frac{1 + \overline{\Delta q_m}^{(0 \to g)}}{n_g\left[1 + \overline{\delta L_m}^{(0 \to g)}\right]}, \quad \text{(SM-3)}$$

respectively. Here, $\Delta q_m^{(0 \to Res)}$ and $\Delta q_m^{(0 \to g)}$ are the number of modes in the measurement cavity the laser jumps from the case with pure vacuum (represented by the 0 in the superscript) to when the cavity contains gas with a refractivity of $n_{Res} - 1$ and $n_g - 1$ (represented by $Res$ and $g$), respectively. Moreover, $\delta L_m^{(0 \to Res)}$ and $\delta L_m^{(0 \to g)}$ represent the physical elongations of the cavity due to the presence of the gas under the same two conditions. In the last steps, $\overline{\Delta q_m}^{(0 \to Res)}$, $\overline{\Delta q_m}^{(0 \to g)}$, $\overline{\delta L_m}^{(0 \to Res)}$, and $\overline{\delta L_m}^{(0 \to g)}$ represent the corresponding relative entities, given by $\Delta q_m^{(0 \to Res)} / q_m^{(0)}$, $\Delta q_m^{(0 \to g)} / q_m^{(0)}$, $\delta L_m^{(0 \to Res)} / L_m^{(0)}$, and $\delta L_m^{(0 \to g)} / L_m^{(0)}$, respectively. These expressions are identical to those for ordinary (unmodulated) single-cavity FPC-based refractometry.

For SCM-GAMOR, the reference cavity is kept at a constant pressure, here, to be applicable to the GEq-GAMOR technique, assumed to be vacuum. Hence, the frequency of the laser that addresses the reference cavity (henceforth referred to as the reference laser), $\nu_r^{(0)}$, is, for the case with SCM-GAMOR, all the time given by

$$\nu_r^{(0)} = \frac{q_r^{(0)} c}{2 L_r^{(0)}}, \tag{SM-4}$$

where $q_r^{(0)}$ and $L_r^{(0)}$ are the mode number addressed and the length of the empty reference cavity, respectively.

This implies that when gas is evacuated from (or filled into) the measurement cavity the measured beat frequency alternates between, for the reference assessment, $f_{(0,Res)}$ (henceforth referred to as the evacuated measurement cavity beat frequency) and, for the filled measurement cavity assessment, $f_{(0,g)}$, (henceforth referred to as the filled measurement cavity beat frequency) which are given by

$$f_{(0,Res)} = \nu_r^{(0)} - \nu_m^{(Res)} = \nu_r^{(0)} - \nu_m^{(0)} \frac{1 + \overline{\Delta q_m}^{(0 \to Res)}}{n_{Res} \left[1 + \overline{\delta L_m}^{(0 \to Res)}\right]} \tag{SM-5}$$

and

$$f_{(0,g)} = \nu_r^{(0)} - \nu_m^{(g)} = \nu_r^{(0)} - \nu_m^{(0)} \frac{1 + \overline{\Delta q_m}^{(0 \to g)}}{n_g \left[1 + \overline{\delta L_m}^{(0 \to g)}\right]}, \tag{SM-6}$$

respectively. The subscripts of the beat frequencies, $_{(0,Res)}$ and $_{(0,g)}$, indicate, as above, the refractivity of the reference and the measurement cavities, respectively, where the 0 indicates that the refractivity in the reference cavity is zero, while the $Res$ and $g$ indicate that the refractivities in the measurement cavity are $n_{Res} - 1$ and $n_g - 1$, respectively.

This implies that the shift in beat frequency when the measurement cavity is filled (or evacuated), $\Delta f_{(0,Res \to g)}$, which, in this case, is solely given by the shift in frequency of the measurement laser, can be expressed as

$$\Delta f_{(0,Res \to g)} = f_{(0,g)} - f_{(0,Res)} = v_m^{(Res)} - v_m^{(g)}$$

$$= v_m^{(0)} \frac{\Delta(n-1)_{(Res \to g)} \left[1 + \overline{\delta L}_m^{(Res \to g)} + 2\overline{\delta L}_m^{(0 \to Res)}\right] - \overline{\Delta q}_m^{(Res \to g)} + \left[1 + (n_{Res} - 1)\right]\overline{\delta L}_m^{(Res \to g)}}{\left[1 + \Delta(n-1)_{(Res \to g)}\right]\left[1 + \overline{\delta L}_m^{(Res \to g)} + 2\overline{\delta L}_m^{(0 \to Res)}\right]}$$

(SM-7)

where we have introduced $\overline{\delta L}_m^{(Res \to g)} = \overline{\delta L}_m^{(0 \to g)} - \overline{\delta L}_m^{(0 \to Res)}$ and $\overline{\Delta q}_m^{(Res \to g)} = \overline{\Delta q}_m^{(0 \to g)} - \overline{\Delta q}_m^{(0 \to Res)}$, and where we have used $\Delta(n-1)_{(Res \to g)}$ as a short hand description of the change in refractivity, i.e. $(n_g - 1) - (n_{Res} - 1)$, which alternatively can be expressed as $n_g - n_{Res}$.

For the case when $(n_{Res} - 1) \ll (n_g - 1)$, which normally is the case for SCM-GAMOR,[1] whereby also $\overline{\delta L}_m^{(0 \to Res)} \ll \overline{\delta L}_m^{(0 \to g)}$,[2] this can be simplified to

$$\Delta f_{(0,Res \to g)} = v_m^{(0)} \frac{\Delta(n-1)_{(Res \to g)} \left[1 + \overline{\delta L}_m^{(Res \to g)}\right] - \overline{\Delta q}_m^{(Res \to g)} + \overline{\delta L}_m^{(Res \to g)}}{\left[1 + \Delta(n-1)_{(Res \to g)}\right]\left[1 + \overline{\delta L}_m^{(Res \to g_g)}\right]}. \qquad (SM-8)$$

This shows that a change in the refractivity in the measurement cavity, from $(n_{Res} - 1)$ to $(n_g - 1)$, will give rise to a change in the frequency of the laser locked to the measurement cavity (possibly exposed to $\Delta q_m^{(Res \to g)}$ mode jumps) and thereby also the beat frequency, of $\Delta f_{(0,Res \to g)}$.

---

[1] For the case with $N_2$, $(n_g - 1)$ is, at atmospheric pressures, ca. $3 \times 10^{-4}$, while $(n_{Res} - 1)$, which often is the refractivity for a pressure of around 1 Pa, frequently is around $3 \times 10^{-9}$.

[2] For the case with $N_2$, $\overline{\delta L}_m^{(0 \to g)}$ is, at atmospheric pressures, often in the $10^{-7} - 10^{-6}$ range, while $\overline{\delta L}_m^{(0 \to Res)}$ often is in the $10^{-11} - 10^{-12}$ range.

Writing $\overline{\delta L}_m^{(0\to g)}$ and $\overline{\delta L}_m^{(0\to Res)}$ as $\varepsilon_m(n_g-1)$ and $\varepsilon_m(n_{Res}-1)$, as in Axner et al.,[1] implies that $\overline{\delta L}_m^{(Res\to g)}$ can be written as $\varepsilon_m \cdot \Delta(n-1)_{(Res\to g)}$. Inserting this into Eq. (SM-8) and solving for $(n_g-1)$ yields

$$n_g - 1 \approx \frac{\overline{\Delta f}_{(0,Res\to g)} + \overline{\Delta q}_m^{(Res\to g)}}{1 - \overline{\Delta f}_{(0,Res\to g)} + \varepsilon_m} + (n_{Res}-1), \qquad \text{(SM-9)}$$

where we have denoted the relative change in beat frequency, $\Delta f_{(0,Res\to g)} / \nu_m^{(0)}$, by $\overline{\Delta f}_{(0,Res\to g)}$. These expressions are identical to those for ordinary (unmodulated) DFPC-based refractometry. The latter one is also identical to Eq. (1) in Silander et al.[2] (neglecting dispersion).

This expression shows that to assess $n_g - 1$ from an measurement of $\overline{\Delta f}_{(0,Res\to g)}$ requires knowledge about the residual refractivity (in the measurement cavity), $n_{Res} - 1$. Since the residual pressure is significantly smaller than that corresponding to $n_g - 1$, it can be assessed with good accuracy by a pressure gauge. From such an assessment, $n_{Res} - 1$ can be calculated. This can then be used to recalculate the evacuated measurement cavity beat frequency, $f_{(0,Res)}$, to its corresponding empty measurement cavity beat frequency $f_{(0,0)}$, which represents the beat frequency when both the reference and the measurement cavities are empty. By this, Eq. (SM-9) can be simplified to

$$n_g - 1 = \frac{\overline{\Delta f}_{(0,0\to g)} + \overline{\Delta q}_m^{(Res\to g)}}{1 - \overline{\Delta f}_{(0,0\to g)} + \varepsilon_m}, \qquad \text{(SM-10)}$$

where $\overline{\Delta f}_{(0,0\to g)}$ is given by $\Delta f_{(0,0\to g)} / \nu_m^{(0)}$, where, in turn, $\Delta f_{(0,0\to g)}$ is given by $f_{(0,g)} - f_{(0,0)}$. This is the expression used for assessment of refractivity in SCM-GAMOR.

The difference between GAMOR and ordinary (unmodulated) DFPC-based refractometry is how the beat frequency for the reference assessment, $f_{(0,Res)}$, and thereby $f_{(0,0)}$, is assessed. Due to drifts in the lengths of the two cavities, these entities will not be true constants, neither through an entire measurement series, nor during a single measurement cycle; they will drift. Ideally, $f_{(0,0)}$ should therefore be assessed at the same time as $f_{(0,g)}$. However, since this is not technically possible, it is, in practice, for ordinary (unmodulated) DFPC-based refractometry, measured at some instance either before or after the filled measurement cavity assessment. This introduces drifts into the system. GAMOR circumvents this problem by estimating $f_{(0,0)}$ at the time instance as $f_{(0,g)}$ is measured by a linear interpolation between two reference assessments (for SMC-GAMOR, two empty measurement cavity assessments) — one performed directly prior to when the measurement cavity is filled with gas and another directly after it has been evacuated — as

$$\tilde{f}_{(0,0)}(t) = f_{(0,0)}(t_n) + \frac{f_{(0,0)}(t_{n+1}) - f_{(0,0)}(t_n)}{t_{n+1} - t_n}(t - t_n), \qquad \text{(SM-11)}$$

where $f_{(0,0)}(t_n)$ and $f_{(0,0)}(t_{n+1})$ represent the reference assessments before the filling of cycle *n* and after the emptying of the same cavity (represented by the assessment performed before the filling of the next cycle, *n+1*). By evaluating this for the same time instance as $f_{(0,g)}$ is measured, the linear drifts in $f_{(0,Res)}$ are eliminated. This implies that the refractivity to be assessed in SCM-GAMOR is evaluated by use of Eq. (SM-10) in which the relative change in beat frequency, $\overline{\Delta f}_{(0,0 \to g)}$, is taken as $[f_{(0,g)} - \tilde{f}_{(0,0)}]/\nu_m^0$, where $\tilde{f}_{(0,0)}$, referred to as the interpolated empty measurement cavity beat frequency, is given by Eq. (SM-11) evaluated at the same time instance as $f_{(0,g)}$ is measured.

**GEq-GAMOR**

For the case when GEq-GAMOR (Gas Equilibration GAMOR) is used, the procedure for the filled measurement cavity assessment is identical to that in SMC-GAMOR, i.e. with the measurement cavity being filled with the gas whose refractivity is to be assessed while the reference cavity is being evacuated. This implies that the frequency of the measurement laser, $v_m^{(g)}$, is given by Eq. (SM-3). However, in this case, the reference cavity will not be fully evacuated (i.e. to vacuum) since it was filled with gas in the previous cycle. This implies that the frequency of the reference laser, which addresses the reference cavity, now denoted $v_r^{(Res)}$, can be expressed as

$$v_r^{(Res)} = \frac{\left[q_r^{(0)} + \Delta q_r^{(0 \to Res)}\right] c}{2 n_{Res} \left[L_r^{(0)} + \delta L_r^{(0 \to Res)}\right]} = v_r^{(0)} \frac{1 + \overline{\Delta q_r}^{(0 \to Res)}}{n_{Res} \left[1 + \overline{\delta L_r}^{(0 \to Res)}\right]}, \qquad \text{(SM-12)}$$

where now $n_{Res}$ is the residual index of refraction in the reference cavity. The filled measurement cavity beat frequency, which now is denoted $f_{(Res,g)}$, is thereby given by $v_r^{(Res)} - v_m^{(g)}$.

For the reference measurements, on the other hand, instead of evacuating the measurement cavity, the two cavities are connected to each other so that the pressures in them are equalized, whereby they will have the same refractivity, denoted $n_{Eq} - 1$. This implies that, during the reference assessments, the frequencies of the two lasers, $v_r^{(Eq)}$ and $v_m^{(Eq)}$, become

$$v_r^{(Eq)} = \frac{\left[q_r^{(0)} + \Delta q_r^{(0 \to Eq)}\right]c}{2n_{Eq}\left[L_r^{(0)} + \delta L_r^{(0 \to Eq)}\right]} = v_r^{(0)} \frac{1 + \overline{\Delta q_r}^{(0 \to Eq)}}{n_{Eq}\left[1 + \overline{\delta L_r}^{(0 \to Eq)}\right]}, \qquad \text{(SM-13)}$$

and

$$v_m^{(Eq)} = \frac{\left[q_m^{(0)} + \Delta q_m^{(0 \to Eq)}\right]c}{2n_{Eq}\left[L_m^{(0)} + \delta L_m^{(0 \to Eq)}\right]} = v_m^{(0)} \frac{1 + \overline{\Delta q_m}^{(0 \to Eq)}}{n_{Eq}\left[1 + \overline{\delta L_m}^{(0 \to Eq)}\right]}, \qquad \text{(SM-14)}$$

where $\Delta q_r^{(0 \to Eq)}$ and $\delta L_r^{(0 \to Eq)}$ are the number of modes the reference laser jumps and the elongation of the length of the reference cavity when it is filled with gas to the equilibration condition, while $\Delta q_m^{(0 \to Eq)}$ and $\delta L_m^{(0 \to Eq)}$ are the corresponding entities for the measurement laser and the measurement cavity, respectively. This implies that the beat frequency measured under gas equilibrated conditions, $f_{(Eq,Eq)}$, henceforth referred to as the gas equilibrium beat frequency, becomes

$$\begin{aligned} f_{(Eq,Eq)} &= v_r^{(Eq)} - v_m^{(Eq)} \\ &= v_r^{(0)} \frac{1 + \overline{\Delta q_r}^{(0 \to Eq)}}{n_{Eq}\left[1 + \overline{\delta L_r}^{(0 \to Eq)}\right]} - v_m^{(0)} \frac{1 + \overline{\Delta q_m}^{(0 \to Eq)}}{n_{Eq}\left[1 + \overline{\delta L_m}^{(0 \to Eq)}\right]}. \end{aligned} \qquad \text{(SM-15)}$$

This expression shows that the beat frequency under gas equilibrium conditions is dissimilar to the evacuated measurement cavity beat frequency measured in SCM-GAMOR when both cavities are evacuated, given by Eq. (SM-5). This implies that also the shift in beat frequency between the filled measurement cavity assessment and the reference assessment is dissimilar to that measured in SCM-GAMOR. Hence, Eq. (SM-10) is not valid for GEq-GAMOR in its present form.

There are a few means to accommodate this. The one used here, which is simple and convenient, is to recalculate which fully evacuated measurement cavity beat frequency (henceforth referred to as the estimated empty measurement cavity beat frequency and denoted $f_{(0,0)}^{(Eq)}$) the assessed gas equilibrium beat frequency, $f_{(Eq,Eq)}$, corresponds to. Such a derivation yields (where we have neglected terms that under normal conditions contribute to the final assessment of refractivity less than $10^{-7}$ on a relative scale, i.e. $< 0.1$ ppm)

$$f_{(0,0)}^{(Eq)} \approx \left[1+\left(n_{Eq}-1\right)\right] f_{(Eq,Eq)} - Q_{Eq} v_m^{(0)} - \Delta\varepsilon(n_{Eq}-1)v_m^{(0)} \qquad \text{(SM-16)}$$

where

$$Q_{Eq} = \overline{\Delta q_m}^{(0 \to g)} - \overline{\Delta q_m}^{(Eq \to g)} - \overline{\Delta q_r}^{(0 \to Eq)} v_r^{(0)} / v_m^{(0)}, \qquad \text{(SM-17)}$$

and where we have expressed $\overline{\delta L_r}^{(0 \to Eq)}$ and $\overline{\delta L_m}^{(0 \to Eq)}$ as $\varepsilon_r(n_{Eq}-1)$ and $\varepsilon_m(n_{Eq}-1)$ and used $\Delta\varepsilon$ for $\varepsilon_m - \varepsilon_r$. We have also eliminated $\Delta q_m^{(0 \to Eq)}$, which in practice is never measured, by using the fact that $\overline{\Delta q_m}^{(0 \to g)} = \overline{\Delta q_m}^{(0 \to Eq)} + \overline{\Delta q_m}^{(Eq \to g)}$.

By doing this, the refractivity of the gas in the measurement cavity during the filled measurement cavity assessment can still be assessed by the use of Eq. (SM-10), although with $\overline{\Delta f}_{(0,0 \to g)}$ being given by $\overline{f}_{(0,g)} - \tilde{\overline{f}}_{(0,0)}^{(Eq)}$, where $\overline{f}_{(0,g)}$ is given by $f_{(0,g)} / v_m^{(0)}$, where in turn, $f_{(0,g)}$ is recalculated from $f_{(Res,g)}$ by the use of the residual refractivity in the reference cavity, $(n_{Res}-1)$. $\tilde{\overline{f}}_{(0,0)}^{(Eq)}$ is given by $\tilde{f}_{(0,0)}^{(Eq)} / v_m^{(0)}$, where the tilde sign indicates that $f_{(0,0)}^{(Eq)}$ should be exposed to an interpolation according to Eq. (SM-11).

**Useful expressions for GAMOR when the refractometer is connected to an external pressure source**

For the case when the pressure in the measurement cavity during the filled measurement cavity assessment is equilibrated with that of an external pressure source (or pressure standard) with a pressure of $P_{Ext}$, with a corresponding refractivity of $n_{Ext} - 1$, Eq. (SM-9) , when used for SCM-GAMOR, can be written as

$$n_{Ext} - 1 \approx \frac{\overline{\Delta f}_{(0, Res \to Ext)} + \overline{\Delta q}_m^{(Res \to Ext)}}{1 - \overline{\Delta f}_{(0, Res \to Ext)} + \varepsilon_m} + (n_{Res} - 1), \qquad (SM\text{-}18)$$

where $\overline{\Delta f}_{(0, Res \to Ext)}$ represents $\Delta f_{(0, Res \to Ext)} / \nu_m^{(0)}$, where $\Delta f_{(0, Res \to Ext)}$ is given by $f_{(0, Ext)} - f_{(0, Res)}$.

Similarly, Eq. (SM-10) can be written as

$$n_{Ext} - 1 = \frac{\overline{\Delta f}_{(0,0 \to Ext)} + \overline{\Delta q}_m^{(Res \to Ext)}}{1 - \overline{\Delta f}_{(0,0 \to Ext)} + \varepsilon_m}, \qquad (SM\text{-}19)$$

where $\overline{\Delta f}_{(0,0 \to Ext)}$ is given by $\Delta f_{(0,0 \to Ext)} / \nu_m^{(0)}$, where, in turn, $\Delta f_{(0,0 \to Ext)}$ is, for SCM-GAMOR, given by $f_{(0, Ext)} - \tilde{f}_{(0,0)}$, while it is, for GEq-GAMOR, given by $f_{(0, Ext)} - \tilde{f}_{(0,0)}^{(Eq)}$.

## NOMENCLATURE

| | |
|---|---|
| $A_R$ and $\tilde{B}_\rho$ | The molecular polarizability and the first density virial coefficient or the gas, respectively |
| $B_p(T)$ | The first pressure virial coefficient |
| $f_{(0,0)}$ | The empty measurement cavity beat frequency, given by the beat frequency between the modes addressed in the reference and the measurement cavities when both the reference and the measurement cavities are empty |
| $\tilde{f}_{(0,0)}(t)$ | The interpolated empty measurement cavity beat frequency, estimated by a linear interpolation from two empty measurement cavity reference measurements performed just prior to, and directly after, the filled measurement cavity assessment, evaluated by Eq. (SM-11) |
| $f_{(0,Ext)}$ | The filled measurement cavity beat frequency, given by the beat frequency between the modes addressed in the reference and the measurement cavities when the reference cavity is empty while the measurement cavity contains the gas to be assessed (with a refractivity of $n_{Ext} - 1$) |
| $f_{(0,Res)}$ | The evacuated measurement cavity beat frequency, given by the beat frequency between the modes addressed in the reference and the measurement cavities when the reference cavity is empty while the measurement cavity is evacuated to a refractivity of $n_{Res} - 1$ |
| $f_{(Res,Ext)}$ | The filled measurement cavity beat frequency, given by the beat frequency between the modes addressed in the reference and the measurement cavities when the reference cavity contains residual gas with a refractivity of $n_{Res} - 1$ while the measurement cavity contains the gas to be assessed (with a presumed refractivity of $n_{Ext} - 1$) |
| $f_{(Eq,Eq)}$ | The gas equilibrium beat frequency, given by the beat frequency between the modes addressed in the reference and the measurement cavities when the both the reference and the measurement cavities contain gas that has a refractivity of $n_{Eq} - 1$ |
| $f_{(0,0)}^{(Eq)}$ | The estimated empty measurement cavity beat frequency, given by an estimate of $f_{(0,0)}$ from an assessment of $f_{(Eq,Eq)}$ by use of Eq. (4). |
| $\tilde{f}_{(0,0)}^{(Eq)}(t)$ | The interpolated estimated empty measurement cavity beat frequency, assessed by a linear interpolation from two estimated empty measurement cavity reference assessments performed just prior to, and directly after, the filled measurement cavity assessment, evaluated by Eq. (SM-11) |
| $\Delta f_{(0,Res \to Ext)}$ | The shift in beat frequency when the measurement cavity is filled from a residual pressure with a refractivity of $n_{Res} - 1$ to a pressure corresponding to a refractivity of $n_{Ext} - 1$, formally given by $f_{(0,Ext)} - f_{(0,Res)}$ |
| $\Delta f_{(0,0 \to Ext)}$ | The shift in beat frequency when the measurement cavity is filled from an empty measurement cavity (i.e. pure vacuum) to a pressure corresponding to a refractivity of $n_{Ext} - 1$, formally given by $f_{(0,Ext)} - f_{(0,0)}$ |
| $\overline{\Delta f}_{(0,0 \to Ext)}$ | Relative value of $\Delta f_{(0,0 \to Ext)}$, defined as $\Delta f_{(0,0 \to Ext)} / q_m^{(0)}$ |
| $\widetilde{\Delta f}_{(0,0 \to Ext)}$ | The shift in beat frequency when the measurement cavity is filled from an empty measurement cavity to a pressure corresponding to a refractivity of $n_{Ext} - 1$ where the reference frequency is estimated by interpolation, formally given by $f_{(0,Ext)} - \tilde{f}_{(0,0)}$ |
| $\Delta f_{(Eq \to Res, Eq \to Ext)}$ | The shift in beat frequency during a GEq-GAMOR cycle between the filled measurement cavity assessment, $f_{(Res,Ext)}$, and the gas equilibrium reference assessment, i.e. when both cavities contain gas with the same refractivity, $n_{Eq}$, i.e. $f_{(Eq,Eq)}$, given by $f_{(Res,Ext)} - f_{(Eq,Eq)}$ |
| $\Delta f_{(0,Res \to Ext)}$ | The shift in beat frequency when the measurement cavity, after being evacuated to a residual refractivity of $n_{Res} - 1$, is filled with the gas to be characterized (assumed to have a refractivity of $n_{Ext} - 1$), formally given by $f_{(0,Ext)} - f_{(0,Res)}$ |
| $\overline{\Delta f}_{(0,Res \to Ext)}$ | Relative value of $\Delta f_{(0,Res \to Ext)}$, defined as $\Delta f_{(0,Res \to Ext)} / v_m^{(0)}$ |
| $f(t)$ | Continuously measured beat frequency |
| $\Delta f(t)$ | The cavity-drift-corrected shift in beat frequency, given by $f(t) - \tilde{f}_{(0,0)}^{(Eq)}(t)$ |
| $L_m^{(0)}$ | The length of the measurement cavity when being empty |
| $L_r^{(0)}$ | The length of the reference cavity when being empty |
| $\overline{\delta L_m}^{(0 \to Eq)}$ | The relative deformation of the length of the measurement cavity when it is filled with gas to a refractivity of $n_{Eq} - 1$ |
| $\overline{\delta L_r}^{(0 \to Eq)}$ | The relative deformation of the length of the reference cavity when it is filled with gas to a refractivity of $n_{Eq} - 1$ |
| $\delta L_m^{(Res \to Ext)}$ | The change in length of the measurement cavity in GEq-GAMOR when it is filled with gas |
| $n_{Ext} - 1$ | The refractivity of the gas in the measurement cavity provided by the external pressure source |
| $n_{Res} - 1$ | The refractivity of the gas (for SCM-GAMOR: in the measurement cavity, for Eqs-GAMOR: in the reference measurement) during evacuation |
| $p$ | The pressure predicted by a fit to a linearity plot |
| $P_{dw}$ | Estimated pressure provided by the piston gauge |
| $P_{Ext}$ | Pressure provided by the external pressure source, or, the pressure in the measurement cavity assessed by refractometry when the measurement cavity is equilibrated with the external pressure source |
| $P_R$ | Pressure in the measurement cavity assessed by the refractometer |
| $\Delta P$ | The difference between the pressure assessed by the refractometer and that provided by the external pressure source, i.e. $P_R - P_{Ext}$ |
| $\Delta q_m^{(Res \to Ext)}$ | The number of modes the measurement laser jumps in GEq-GAMOR when the refractivity of the measurement cavity is changed from the residual refractivity, $n_{Res} - 1$, to $n_{Ext} - 1$ |
| $\overline{\Delta q_m}^{(Res \to Ext)}$ | The relative value of $\Delta q_m^{(Res \to Ext)}$, defined as $\Delta q_m^{(Res \to Ext)} / q_m^{(0)}$ |
| $\varepsilon_m$ | Refractivity-normalized deformation coefficient of the measurement cavity, defined as $\varepsilon_m (n_{Ext} - n_{Res}) = \delta L_m^{(Res \to Ext)} / L_m^{(0)}$ |
| $\varepsilon_r$ | Refractivity-normalized deformation coefficient of the reference cavity, defined as $\varepsilon_r (n_{Eq} - 1) = \delta L_r^{(0 \to Eq)} / L_r^{(0)}$ |
| $\Delta \varepsilon$ | Shorthand notation for $\varepsilon_m - \varepsilon_r$ |
| $v_m^{(0)}$ | The frequency of the $q_m^{(0)}$:th cavity mode addressed by the laser in the empty measurement cavity |
| $v_r^{(0)}$ | The frequency of the $q_r^{(0)}$:th cavity mode addressed by the laser in the empty reference cavity |
| $\rho_{Ext}$ | Assessed density of the gas in the measurement cavity supplied by the external pressure source |
| Filled measurement cavity assessment: | Assessment of $f_{(0,Ext)}$ |
| Reference assessment: | In SCM-GAMOR, either the evacuated measurement cavity reference assessment, $f_{(0,Res)}$, or the empty measurement cavity reference assessment, $f_{(0,0)}$. |
| | In GEq-GAMOR, either the gas equilibrium reference assessment, $f_{(Eq,Eq)}$, or the estimated empty measurement cavity reference assessment, $f_{(0,0)}^{(Eq)}$. |